\documentclass{jfm}

\usepackage{amsmath,amssymb,mathtools,graphicx,xcolor}

\newcommand{\hinf}{\hat{h}_\infty}
\newcommand{\ii}{\mathrm{i}}
\newcommand{\ee}{\mathrm{e}}
\newcommand{\const}{\mathrm{const}}

\newcommand{\init}{\mathrm{init}}
\newcommand{\scF}{\mathcal{F}}

\newcommand{\scM}{\mathcal{M}}

\newcommand{\scX}{\mathcal{X}}

\newcommand{\scQh}{\hat{\mathcal{Q}}}

\newcommand{\sep}{\qquad}

\DeclareMathAlphabet{\mathbfsf}{\encodingdefault}{\sfdefault}{bx}{sl}
\newcommand{\vect}[1]{\boldsymbol{#1}}

\newcommand{\grad}{\boldsymbol{\nabla}}

\newcommand{\pdiff}[2]{\frac{\partial{#1}}{\partial{#2}}}
\newcommand{\tdiff}[2]{\frac{\mathrm{d}{#1}}{\mathrm{d}{#2}}}

\newcommand{\dotProd}{\boldsymbol{\cdot}}

\makeatletter

\newcommand*\wthelper[2]{\hbox{\dimen@\accentfontxheight#1\accentfontxheight#11.2\dimen@$\m@th#1\widetilde{#2}$\accentfontxheight#1\dimen@}}

\newcommand*\dimlesshelper[2]{\hbox{\dimen@\accentfontxheight#1\accentfontxheight#11.2\dimen@$\m@th#1\widehat{#2}$\accentfontxheight#1\dimen@}}
\newcommand*\accentfontxheight[1]{\fontdimen5\ifx#1\displaystyle
	\textfont
	\else\ifx#1\textstyle
	\textfont
	\else\ifx#1\scriptstyle
	\scriptfont
	\else
	\scriptscriptfont
	\fi\fi\fi3
}
\makeatother

\makeatletter
\newcommand{\customlabel}[2]{%
\protected@write \@auxout {}{\string \newlabel {#1}{{#2}{}}}}
\makeatother

\renewcommand{\vect}[1]{{\bf #1}}
\renewcommand\hinf{h_0}
\renewcommand\scQh{Q}
\newcommand\mua{\mu_a} 
\newcommand\mui{\mu_i}
\newcommand\amp{A}
\newcommand\scN{{\cal N}}
\newcommand\dimp{\tilde p} 
\newcommand\dimq{\tilde {\bf q}}

\title{Fingering instability of self-similar radial flow of miscible fluids in a Hele-Shaw cell}
\shorttitle{Fingering of self-similar Hele-Shaw flow}
\shortauthor{J. R. Lister and T.-F. Dauck}

\author{John R. Lister\aff{1}\corresp{\email{lister@damtp.cam.ac.uk}} \and Tim-Frederik Dauck\aff{1}}

\affiliation{\aff{1}Department of Applied Mathematics and Theoretical Physics, University of Cambridge, Wilberforce Road, Cambridge CB3 0WA, UK}

\begin{document}

\maketitle

\begin{abstract}
The linear stability of miscible displacement for radial source flow at infinite P\'eclet number in a Hele-Shaw cell is calculated theoretically. The axisymmetric self-similar flow is shown to be unstable to viscous fingering if the viscosity ratio $m$ between ambient and injected fluids exceeds $3\over2$ and to be stable if $m<{3\over2}$. If $1<m<{3\over2}$ small disturbances decay at rates between $t^{-3/4}$ and $t^{-1}$ (the exact range depending on $m$) relative to the $t^{1/2}$ radius of the axisymmetric base-state similarity solution; if $m<1$ they decay faster than $t^{-1}$. Asymptotic analysis confirms these results and gives physical insight into various features of the numerically determined relationship between the growth rate and the azimuthal wavenumber and viscosity ratio. 
\end{abstract}


\section{Introduction}\label{sec:intro}

Viscous, or Saffman--Taylor, fingering is one of the canonical fluid-mechanical instabilities, which occurs when a lower viscosity fluid is driven into a restricted environment occupied by fluid with a (sufficiently) higher viscosity. In the context of flow in a porous medium \citep[e.g.][]{Homsy1987}, it has huge economic impact by significantly reducing the efficiency of oil extraction from reservoirs by water injection \citep{Lake}. In the context of flow in a Hele-Shaw cell, it has generated significant scientific interest following the seminal paper by \citet{Saffman1958} as a prototypical example of pattern formation and selection, particularly in the limit of small or vanishing surface tension \citep[see, e.g.][]{Couder2000,Bischofberger2014,Andersen2024}. Other recent work considers the possible suppression or modification of Saffman--Taylor fingering by varying the geometry of the Hele-Shaw cell boundaries \citep[e.g.][]{PihlerPuzovic2012, AlHousseiny2012,Zheng2015, Peng2019}, viscous fingering in two-layer viscous gravity currents \citep[e.g.][]{Kowal2019a,Kowal2019b,Dauck} and bubble compressibility \citep{Cuttle2023}.

The linear instability of {\it immiscible} displacement in a Hele-Shaw cell or porous medium has been variously analysed for both unidirectional and radial motion \citep[e.g.][]{Hill1952,Saffman1958,Chouke1959,Wilson1975,Paterson1981}. For simplicity, it is assumed that the intruding fluid displaces all of the ambient fluid, though several authors comment that the analysis is easily adapted to the case where a constant-thickness layer of ambient fluid is left behind \citep[cf.][]{Park1984} and coats the cell walls. In the simple case, the jump in the viscous pressure gradient at the interface drives growth of interfacial perturbations at a rate proportional to $(\mua-\mui)kV/(\mua+\mui)$, where $\mua$ is the ambient viscosity,  $\mu_i$ the intruding fluid viscosity, $k$ the transverse wavenumber and $V$ the displacement velocity. Gravity, surface tension and a radial geometry may provide stabilising effects, but if the viscosity ratio $m\equiv \mua/\mui$ exceeds $1$, and $V$ is sufficiently large, the flow will be unstable to what has become known as Saffman--Taylor fingering.
Surface tension does stabilise the short wavelengths, leading to a most-unstable wavelength proportional to, and a growth rate inversely proportional to, the square root of the surface tension. Hence the limit of zero surface tension appears singular, but it can be regularised \citep[e.g.][]{Paterson1985,Dias2013,Nagel2013} by re-including finite-aspect-ratio effects, which lead to a most-unstable wavelength comparable to the cell thickness in accord with observations.

The case of zero surface tension also arises naturally when considering {\it miscible} displacement in a Hele-Shaw cell. \citet{Wooding1969} first described experimentally viscous fingering in a Hele-Shaw cell with miscible fluids similar to that seen with immiscible fluids. \citet{Paterson1985} presented a stability analysis for an inviscid intrusion spreading from a point source into a miscible viscous ambient with diffusion assumed negligible. Significantly, Paterson neglected radial variations in the thickness of the intrusion by assuming that the ambient leaves at most a thin and immobile film of constant thickness behind the front. His model is then effectively equivalent to the case of complete displacement with immiscible fluids in the zero-surface-tension limit. However, if the intrusion fluid is viscous then miscible displacement results in an intruding tongue of fluid along the centre of the channel whose thickness varies with radial position \citep{Petitjeans1996,Chen1996,Rakotomalala1997,Yang1997}. The resultant viscosity variation across the cell causes the velocity profile to differ from the simple Poiseuille profile of immiscible displacement (see figure \ref{fig:setup}). 
Moreover, at low or moderate P\'eclet number the radial and vertical viscosity structure of miscible displacement is also affected by cross-flow diffusion and radial dispersion \citep[e.g.][]{Tan1987,Goyal2006,Nijjer2018,Videbaek2019,Sharma2020}, and even the unperturbed base state is time-dependent and must be determined numerically. At large P\'eclet number, however, it is reasonable to neglect diffusion until the dimensionless radius is comparably large, and it is possible to make more progress analytically.

\citet{Yang1997} analysed unidirectional miscible displacement with negligible diffusion and obtained a kinematic-wave equation for the height of the intruding tongue of fluid. For  viscosity ratios $m<{3\over2}$ they found a smooth similarity solution (with no shocks) as a function of $x/t$. However, for $m>{3\over2}$ the kinematic-wave equation necessarily forms a frontal shock of a height that they recognised, in principle, might require a fully two-dimensional Stokes-flow calculation near the nose to determine \citep[cf.][]{Goyal2006}. 
As discussed further in \S3.1, they instead presented a classical tangent construction of a so-called `contact' shock from the flux function of the intruding fluid. 
In terms of the intruding fluid fraction $\lambda_*$ at the nose, the contact-shock height is given by $\lambda_*=\lambda_c(m)$, where 
\begin{equation}\label{lambdaCrit}
	\lambda_c=
		2\left(\tfrac{2}{3}m-1\right)^{-1/2}\sinh\left[\tfrac{1}{3}\sinh^{-1}\left\{(m-1)^{-1}\left(\tfrac{2}{3}m-1\right)^{3/2}\right\}\right],  \quad (m>\tfrac{3}{2}) \,,
\end{equation}
is the real root of a certain cubic polynomial.
They note that experiments and numerical models \citep{Petitjeans1996,Chen1996,Rakotomalala1997} suggest that (\ref{lambdaCrit}) underestimates the shock height, particularly for $m>5$, and more recent experiments \citep{Bischofberger2014,Videbaek2020} support this. Limited data for smaller $m$ is roughly consistent with (\ref{lambdaCrit}). To the best of our knowledge, the exact nature of these shocks and how their height is determined is not yet fully understood, perhaps because they are experimentally unstable to fingering in the transverse direction. 

\citet{Lajeunesse1997,Lajeunesse1999,Lajeunesse2001} and later \citet{Bischofberger2014} conducted experiments in Hele-Shaw cells with miscible fluids and negligible diffusion, and observed a viscous fingering instability at the front of the intrusion if $m$ was sufficiently large. The minimum viscosity ratio for instability to be seen was 2--3, a little larger than the critical value $m={3\over2}$ derived by \citet{Yang1997}, and clearly larger than the critical value $m=1$ for the case of immiscible displacement. Both \citet{Lajeunesse1997,Lajeunesse1999,Lajeunesse2001} and \citet{Bischofberger2014} suggested that the existence of a flat shock front and the associated jump in pressure gradient across the front are crucial to the development of a fingering instability similar to the classical Saffman--Taylor instability. \citet{Lajeunesse2001} approximated the intrusion as being of uniform thickness equal to the shock height and, by adapting Saffman \& Taylor's analysis appropriately, obtained a good prediction of the instability threshold for vertical displacement. Recently, \citet{Videbaek2020} adopted the same approach to obtain a good prediction for the instability onset radius for radial flow. (Videbaek also provides an interesting synthesis of experimental observations of immiscible and miscible displacements in linear and radial geometries.)

From the preceding work, it seems to be accepted that $m={3\over2}$ is the predicted stability boundary for miscible intrusion without diffusion. However, to the best of our knowledge,
it has not been demonstrated theoretically that the flow is stable for $m<{3\over2}$ and it has been difficult to demonstrate experimentally that the flow is unstable for $m$ less than about 2--3 \citep{Lajeunesse1997,Bischofberger2014,Videbaek2020}. Indeed \citet{Bischofberger2014} note ``It is important to point out that both the connection between the shock-front formation and the onset of the lateral instability, and the suppression of any instability (for example, of the kind from the original Saffman--Taylor analysis) for $0.67<\eta_{in}/\eta_{out}<1$ [i.e.~$1<m<{3\over2}$] remain to be explained.'' It is our intention to provide some explanation in this paper.

We consider the linear stability of miscible displacement with negligible diffusion from a point source in Hele-Shaw flow, which is parameterised by the viscosity ratio $m$ between the ambient and intruding fluids. The set-up of the mathematical model, its governing equations, assumptions and nondimensionalisation are described in \S\ref{sec:setup}. An analytic solution to the initial-value problem is found in \S\ref{sec:kwave} for the special case of axisymmetric flow  using the method of characteristics. In the absence of any nonaxisymmetric perturbations, this kinematic-wave solution tends towards a simple axisymmetric similarity solution like $t^{-1}$. The central point of the paper is a linear stability analysis of this similarity solution in \S\ref{sec:linearstability}.  Working in similarity space, in \S\ref{sec:linearstability-eqns} we derive coupled ordinary differential equations for the radial structure of eigenmodes of specified azimuthal wavenumber and in \S\ref{sec:linearstability-results} we present numerical results for their growth rates as functions of wavenumber and viscosity ratio. Further insight into the structure of the problem and some good asymptotic approximations to the numerical results are obtained in \S\ref{sec:largek} and in Appendix \ref{AppB} by analysing the various modes using the WKB (Wentzel--Kramers--Brillouin) method for large azimuthal wavenumber. We confirm stability for $m<{3\over2}$ and explain why instability is hard to observe for $m$ only somewhat larger than ${3\over2}$. We  discuss our conclusions in \S\ref{discussion}.

\section{Model description}\label{sec:setup}
\subsection{Governing equations}\label{sec:setup-eqns}

\begin{figure}
	\centering
	\includegraphics[width=0.85\textwidth]{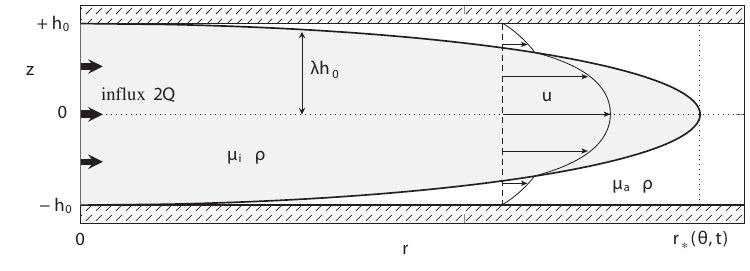}
	\caption{A radial cross-section of an axisymmetric intrusion with constant influx $2\scQh$ into a Hele-Shaw cell with gap thickness $2\hinf$. The shape of the intrusion is described by the intruding fluid fraction $\lambda(r,\theta,t)$ and its radial extent $r_*(\theta,t)$. The viscosities, $\mui$ and $\mua$, of the intruding and ambient fluids give rise to a viscosity ratio $m=\mua/\mui$. The densities $\rho$ are equal. The velocity profile is piecewise parabolic and given by (\ref{velocity}).
	}
	\label{fig:setup}
\end{figure}

Consider radial flow from a point source in a Hele-Shaw cell consisting of  infinite parallel rigid plates separated by a constant distance $2\hinf$. The cell is initially filled with ambient fluid of viscosity $\mua$ and density $\rho$. For $t>0$, fluid of viscosity $\mui$ and (equal) density $\rho$ is injected at the origin at a constant volume flux $2\scQh$.  As shown in figure~\ref{fig:setup}, we use cylindrical polar coordinates $(r,\theta,z)$ to describe the horizontal extent of the current $r_{*}(\theta,t)$ and the vertical thickness of the intrusion $2\hinf\lambda(r,\theta,t)$, where $\lambda(r,\theta,t)$ is the local fluid fraction of injected fluid. Surface tension, diffusion and inertia are all assumed to be negligible.

After an initial transient, the horizontal extent of the intrusion is much greater than its vertical extent, $r_*\gg \hinf$. In this limit, the vertical velocity is negligible and the horizontal velocity $\vect{u}(r,z,t)$ is related to the horizontal pressure gradient $\grad\dimp$ by the lubrication approximation 
\begin{equation}
    \label{lubrication}
	\mu\pdiff{^2\vect{u}}{z^2}=\grad\dimp  
\end{equation}
subject to boundary conditions
\begin{equation}\label{velocityBC}		
	\vect{u}=\vect{0}\text{ at }z=\pm \hinf,\qquad
	\left[\vect{u}\right]_-^+=\vect{0}\text{ and }
	\left[\mu\pdiff{\vect{u}}{z}\right]_-^+=\vect{0}\text{ at }z=\pm \lambda\hinf,
\end{equation}
which impose no-slip on the boundaries, and velocity and stress continuity at the interfaces between the fluids. Solution of (\ref{lubrication}) and (\ref{velocityBC}) yields the velocity profile
\begin{subequations}\label{velocity}\begin{align}
	\hspace{30pt}
	\vect{u}&=\frac{\grad\dimp  }{2\mua}\left(mz^2-\hinf^2\{1+(m-1)\lambda^2\}\right)
	&
	\text{for}&&|z|<\lambda\hinf,
	\hspace{30pt}
	\\
	\vect{u}&=\frac{\grad\dimp  }{2\mua}\left(z^2-\hinf^2\right)
	&
	\text{for}&&\lambda\hinf<|z|<\hinf,
	\hspace{30pt}
\end{align}\end{subequations}
whose shape depends on the viscosity ratio $m=\mua/\mui$ and the intruding fluid fraction $\lambda$. Integration of (\ref{velocity}{\it a}) between $\pm\lambda\hinf$ gives the horizontal flux $2\dimq_i$ of intruding fluid, while integration of (\ref{velocity}) between $\pm\hinf$ gives the total flux $2\dimq$. We obtain
\begin{subequations}\label{dimq}
\begin{align}
	2\dimq_i&=
		-\frac{\hinf^3}{3\mua} \grad\dimp \left\{3\lambda+(2m-3)\lambda^3\right\},
		\label{dimqi}\\
	2\dimq&=
		-\frac{2\hinf^3}{3\mua}\grad\dimp  \left\{1+(m-1)\lambda^3\right\}.
		\label{dimqt}
\end{align}
\end{subequations}

Using these fluxes, we can straightforwardly obtain two local mass-conservation equations for $r<r_*(\theta,t)$:
\begin{subequations}\label{dimPDE}
\begin{equation}
	\pdiff{\lambda}{t}=
		\frac{\hinf^2}{6\mua}\grad\dotProd\left(\grad\dimp  \left\{3\lambda+(2m-3)\lambda^3\right\}\right),
		\label{dimPDE-interface}
\end{equation}
\begin{equation}
		\grad\dotProd\left(\grad\dimp  \left\{1+(m-1)\lambda^3\right\}\right)=0.
		\label{dimPDE-pressure}
\end{equation}
Equation (\ref{dimPDE-interface}) determines the evolution of $\lambda$ from conservation of  intruding fluid, while (\ref{dimPDE-pressure}) determines the pressure gradient $\grad\dimp  $ from a divergence-free constraint on the total flux due to the fixed cell boundaries.
Ahead of the intrusion, in $r>r_*(\theta,t)$, we have $\lambda=0$ and thus
(\ref{dimPDE-interface}) is not relevant and (\ref{dimPDE-pressure}) reduces to 
\begin{equation}\label{dimPDE-laplace}
	\nabla^2\dimp  =0.
\end{equation}\end{subequations}
We assume there is no imposed far-field pressure gradient, i.e.\ $\grad\dimp  \to0$ as $r\to\infty$ and, in the absence of surface tension, there is no capillary pressure jump at the nose of the intrusion.

The injection of intruding fluid (only) at the origin at a constant flux $2\scQh$ corresponds to the boundary conditions that $\lambda=1$ at $r=0$ and 
\begin{equation}\label{dimBC-globalMass}
	-\frac{m\hinf^3}{3\mua}\lim_{r\to0}\left(2\pi r\,\vect{e}_r\dotProd\grad\dimp  \right) = \scQh,
\end{equation}
where $\vect{e}_r$ is the radially outward unit vector. 

At $r=r_*(\theta,t)$, continuity of pressure and  of the total flux normal to the nose yields
\begin{align}\label{dimBC-cts}
	\left[\dimp  \right]_-^+&=0,
	\sep
	\left[\frac{\hinf^2\left(\vect{n}\dotProd\grad\dimp  \right)}{3\mua}\left\{1+(m-1)\lambda^3\right\}\right]_-^+=0
	\sep\text{at}\sep r=r_*,
\end{align}
where $\vect{n}$ is the normal to the perimeter $r=r_*(\theta,t$) of the intrusion. The flux of intruding fluid normal to the nose also has to be consistent with the normal velocity of the nose, which gives
\begin{equation}\label{dimBC-noseMass}
	\left(\vect{n}\dotProd\vect{e}_r\right)\pdiff{r_*}{t}=\frac{\hinf^2\left(\vect{n}\dotProd\grad\dimp  \right)}{6\mua}\left\{3+(2m-3)\lambda_*^2\right\},
\end{equation}
where $\lambda_*$ is the limiting value of $\lambda$ as $r\to r_{*-}$. For the case of a rounded nose with $\lambda_*=0$, (\ref{dimBC-noseMass}) is just the kinematic condition that the nose moves with the centreline velocity. For the case of a frontal shock with $\lambda_*>0$, (\ref{dimBC-noseMass}) is the condition of mass conservation of intruding fluid across the shock.

Equations (\ref{dimPDE})--(\ref{dimBC-noseMass}), together with a condition such as (\ref{lambdaCrit}) on any frontal shock height, describe the evolution of the spreading intrusion in terms of the dimensional pressure $\dimp$, the intruding fluid fraction $\lambda$ and the shape of the perimeter $r_*$. 
These equations for non-axisymmetric flow are equivalent to those of \citet{Yang1997} for unidirectional flow.

\subsection{Non-dimensionalisation and similarity variables}\label{sec:setup-nondim}

The intrusion volume suggests a rough scaling $r_*^2\hinf\sim Qt$ and (\ref{dimBC-noseMass}) suggests $r^*/t\sim \hinf^2\,\dimp /r^* \mua$. More detailed scaling of (\ref{dimPDE-interface}) and (\ref{dimBC-globalMass}) provides numerical factors and motivates definition of a radial similarity variable $\xi$ and a dimensionless pressure $p$ by
\begin{equation}\label{eqn:setup-similarityVars}
	\xi=\left(\frac{2\pi\hinf}{\scQh}\right)^{1/2}\frac{r}{t^{1/2}}
	\qquad\text{and}\qquad
	p(\xi,\theta,t)=\left(\frac{2\pi\hinf^3}{3\mua\scQh}\right)\dimp  (r,\theta,t).
\end{equation}
We also define a mobility function $\scM$ and a flux function $\scF$ by
\begin{equation}\label{MFdef}
	\scM(\lambda;m)=1+(m-1)\lambda^3
	\qquad\text{and}\qquad
	\scF(\lambda;m)=\frac{3\lambda+(2m-3)\lambda^3}{2+2(m-1)\lambda^3},
\end{equation} 
which give the relative mobility for the total flux and the flux fraction of intruding fluid respectively.
As usual for description of evolution towards self-similarity \citep[e.g.][]{Witelski1999,Leppinen2003,Mathunjwa2006,Peng2014}, we define a dimensionless time variable by $\tau=\ln(t/\hat{t})$, where $\hat{t}$ is a reference time scale such as $Q/\hinf^3$.

In terms of the new dimensionless variables, the local mass-conservation equations (\ref{dimPDE}) can be written as the coupled partial differential equations
\begin{subequations}\label{nondimPDE}
\begin{equation}\tag{\ref{nondimPDE}{\it a--c}}
  \vect{q}=-\scM\grad p, \qquad
  \grad\dotProd\vect{q}=0, \qquad
  \pdiff{\lambda}{\tau}-\frac{\xi}{2}\pdiff{\lambda}{\xi}=
          -\grad\dotProd\left(\scF\vect{q}\right)\qquad\text{for}\qquad \xi<\xi_*
\end{equation}
\begin{equation}\tag{\ref{nondimPDE}{\it d,e}}
		\vect{q}=-\grad p,\qquad
		\grad^2p=0 \qquad\text{for}\qquad \xi>\xi_*\,,
\end{equation}
\end{subequations}
where $2\vect{q}(\xi,\theta,\tau)$ is the total flux and $\grad=\vect{e}_\xi\,\partial/\partial\xi +\vect{e}_\theta\, \xi^{-1}\partial/\partial\theta$  now denotes the horizontal gradient operator in similarity space.
The boundary conditions (\ref{dimBC-globalMass})--(\ref{dimBC-noseMass}) can be written as 
\begin{subequations}\label{nondimBCs}
\begin{equation}\tag{\ref{nondimBCs}{\it a,b}}
		\vect{q}\to\xi^{-1}\vect{e}_\xi
		\quad\text{as}\quad\xi\to0, \customlabel{eqn:BC_influx}{\ref{nondimBCs}{\it a}}
		\qquad\qquad
		\vect{q}\to\vect{0}
		\quad\text{as}\quad\xi\to\infty, \customlabel{eqn:BC_infty}{\ref{nondimBCs}b}
\end{equation}
\begin{equation}\tag{\ref{nondimBCs}{\it c--e}}
		[p]^+_-=0, \customlabel{eqn:BC_p_cts}{\ref{nondimBCs}c}
\qquad
		\left[\vect{n}\dotProd\vect{q}\right]^+_-=0
				\customlabel{eqn:BC_q_cts}{\ref{nondimBCs}d}
\quad\text{and}\quad
		\pdiff{\xi_*}{\tau}=\frac{\vect{n}\dotProd\vect{q}}{\vect{n}
		\dotProd\vect{e}_\xi}\frac{\scF_*}{\lambda_*}-\frac{\xi}{2}
        \qquad\text{at}\qquad\xi=\xi_*, 
		\customlabel{eqn:BC_noflux}{\ref{nondimBCs}e}
\end{equation}\end{subequations}
where $\partial\xi_*/\partial\tau$ is the dimensionless speed of the nose in similarity space. 

If the evolution of the system becomes self-similar and independent of $\tau$ at late times, then (\ref{nondimPDE}) reduces to a system of coupled ordinary differential equations.

\section{Axisymmetric flows and similarity solutions}\label{sec:kwave}\label{sec:kwave-axisym}
If the flow is axisymmetric, i.e.\ $\partial/\partial\theta=0$, then $q_\theta=0$ and so $\grad\dotProd\vect{q}=0$ gives $\vect{q}=\xi^{-1}\vect{e}_\xi$ everywhere. Therefore, (\ref{nondimPDE}$c$) becomes
\begin{align}\label{kPDE}
	\pdiff{\lambda}{\tau}=\left(\frac{\xi}{2}-\frac{\scF'(\lambda)}{\xi}\right)\pdiff{\lambda}{\xi},
\end{align}
where $\scF'(\lambda)$ denotes $\partial\scF/\partial\lambda$. This equation is a simple quasilinear hyperbolic equation, which, in the absence of shocks (discontinuities in $\lambda$), can be solved analytically by the method of characteristics. (A similar kinematic-wave construction is given by \cite{Yang1997} and \cite{Lajeunesse1999} for related unidirectional flows.)

Equation (\ref{kPDE}) implies that $\lambda$ is constant along characteristic curves $\xi(\tau)$ defined by
\begin{equation}\label{characteristics}
	\tdiff{\xi}{\tau}=\frac{\scF'(\lambda)}{\xi}-\frac{\xi}{2}
	\sep
	\Longleftrightarrow
	\sep
	\left(\xi^2-2\scF'\right)\ee^{\tau}=\const.
\end{equation}
Thus $\lambda$ maintains its initial value $\lambda(\xi_\init,0)$ on the characteristic that passes through $\xi=\xi_\init$ at $\tau=0$.  Solving (\ref{characteristics}) for $\xi_\init(\xi,\tau)$ thus leads to a solution of (\ref{kPDE}) in the form
\begin{equation}\label{solLambda}
	\lambda(\xi,\tau)=\lambda\left(\left\{\xi^2\ee^\tau+2\scF'\left(1-\ee^\tau\right)\right\}^{1/2},0\right).
\end{equation}
This is implicit in $\lambda$ as the value of $\xi_\init(\xi,\tau)$   depends on $\lambda$ through $\scF'$ and, in general, it is not possible to solve (\ref{solLambda}) for $\lambda(\xi,\tau)$ explicitly. If $\lambda$ varies monotonically with $\xi$ in some region it is, however, possible to exploit a change of variable from $\lambda(\xi,\tau)$ to $\xi(\lambda,\tau)$ to obtain an 
explicit solution for $\xi(\lambda,\tau)$:
\begin{align}\label{solXi}
	\xi(\lambda,\tau)=\left\{\xi_\init^2(\lambda)\ee^{-\tau}+2\scF'\left(1-\ee^{-\tau}\right)\right\}^{1/2}.
\end{align}
We will see later that the fluid fraction $\lambda$ is often a more convenient independent variable than the radial distance $\xi$. 

Provided the characteristics do not cross (no shocks form), (\ref{solXi}) shows that 
\begin{equation}\label{solSteady}
	\xi(\lambda,\tau)\to X_0(\lambda)\equiv\big\{2\scF'(\lambda)\big\}^{1/2}
	\quad\text{as}\quad \tau\to\infty.
\end{equation}
For $m\leq{3\over2}$, $\scF'(\lambda)$ is a monotonically decreasing function for $\lambda\in[0,1]$ and (\ref{solSteady}) describes the shape $X_0(\lambda)$ of a long-time similarity solution in which $\lambda$ varies smoothly from $\lambda=1$ at $\xi=0$ to $\lambda=0$ at $\xi=X_{0*}=\{2\scF'(0)\}^{1/2}$. For $m>{3\over2}$, $\scF'(\lambda)$ is an increasing function for $\lambda$ in a certain range $[0,\lambda_m]$, where $\lambda_m>0$ and $\scF''(\lambda_m)=0$; hence a frontal shock must form by some characteristics for some $\lambda\in(0,\lambda_m)$ overtaking the characteristic for $\lambda=0$, as discussed further below. Neverthless, (\ref{solSteady}) still gives the shape of a long-time similarity solution for $\lambda\in[\lambda_*,1]$, where $\lambda_*$ is the frontal shock height. Similar results were obtained by \citet{Yang1997} for unidirectional flow.

Importantly, as we are interested in the linear stability of radial intrusions into a Hele-Shaw cell, we can expand (\ref{solXi}) as $\tau\to\infty$, to obtain
\begin{equation}\label{solExpansion}
	\xi\sim X_0+\frac{\xi_\init^2-X_0^2}{2X_0}\ee^{-\tau}+\cdots.
\end{equation}
A key implication of (\ref{solExpansion}) is that any axisymmetric perturbations left over from the initial conditions  decay as $O\left(\ee^{-\tau}\right)$ or, equivalently, $O(t^{-1})$. The decay of all axisymmetric perturbations at the same rate in this problem may be contrasted with perturbations from self-similarity in other problems \citep[see, e.g.,][]{Witelski1999,Leppinen2003,Mathunjwa2006} where there is a discrete spectrum of distinct eigenmodes with different decay rates.

\subsection{Solutions with shocks}\label{sec:shocks}

\begin{figure}
	\centering
	\includegraphics[width=0.7\textwidth]{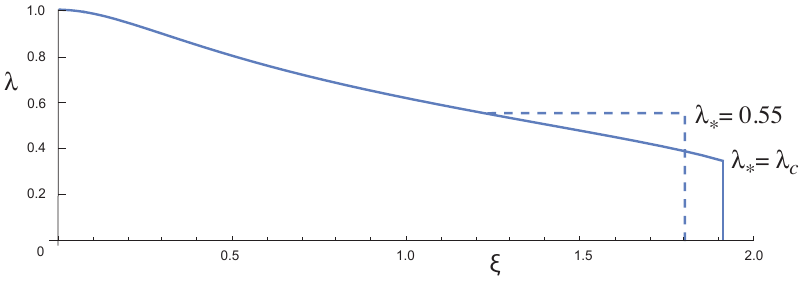}
	\caption{Two possible axisymmetric similarity solutions with different frontal shock heights for an intrusion with viscosity ratio $m=10$. The curved profile for $\lambda>\lambda_*$ is given by (\ref{solSteady}) in both cases. The minimal shock height is $\lambda_*=\lambda_c\approx0.34$ for a contact shock (solid line). Also shown is a possible undercompressive shock of height $\lambda_*=0.55$ (dashed), which travels faster than the characteristics with $\lambda>0.55$, but slower than a contact shock.}
	\label{fig:shocks}
\end{figure}

For $m>{3\over2}$, $\scF'(\lambda)$ does not vary monotonically with $\lambda$, as noted above, and so some characteristics overtake others, which leads to interfacial steepening and shock formation --- in physical terms, we interpret shocks as regions where $\lambda(r)$ varies significantly on the short length scale $\hinf$ rather than the long length scale $r_*$. (An interpretation that the interface folds over to give a multivalued solution for $\lambda$ is unphysical  as the flow profile (\ref{velocity}) decreases monotonically away from the maximum velocity on the centreline.)  Depending on initial conditions, the shock may initially form in the interior of the flow \citep{Dauck2019}, but it will eventually overtake the front and become a frontal shock of some height $\lambda_*$. For axisymmetric flow, the frontal condition (\ref{nondimBCs}$e$) reduces to 
\begin{equation}\label{shockvel}
	\tdiff{\xi_*}{\tau}=\frac{\scF(\lambda_*)}{\lambda_*\xi_*}-\frac{\xi_*}{2}.
\end{equation}
If $\scF(\lambda_*)/\lambda_*<\scF'(\lambda_*)$ then characteristics continue to overtake the front and the shock height increases until $\scF(\lambda_*)/\lambda_*\ge\scF'(\lambda_*)$. There are two cases to consider (see figure \ref{fig:shocks}), which previous work suggests may be relevant for smaller and larger values of $m$ respectively.

If the shock height is determined by the information reaching it along characteristics, then $\lambda_*$ will tend towards the equilibrium height $\lambda_c$ of a so-called `contact' shock, where $\scF(\lambda_c)/\lambda_c=\scF'(\lambda_c)$, with $\lambda_c$ given by (\ref{lambdaCrit}). A consequence of this condition is that $\{\scF(\lambda_c)/\lambda_c\}'=0$ and so the shock speed $\scF(\lambda_*)/\lambda_*$ for small perturbations differs from the equilibrium value only at $O\big((\lambda_*-\lambda_c)^2 \big)$. This does not affect the linear behaviour and hence the shock position $\xi_*$ tends towards its equilibrium position as $\ee^{-\tau}$ just like the characteristics (compare equations \ref{characteristics} and \ref{shockvel}).  

Alternatively, as seems to be the case for at least $m>5$, the shock height is determined by local dynamics on the length scale $\hinf$ of two-dimensional Stokes flow around the front of the intrusion \citep{Yang1997}.
In this case we have $\lambda_*>\lambda_c$, $\xi_*<\xi_c$ and $\scF(\lambda_*)/\lambda_*>\scF'(\lambda_*)$, and the so-called `undercompressive' shock \citep[see][]{Bertozzi1999} outpaces the characteristics to leave a flat region behind it where $\lambda=\lambda_*$ (dashed line in figure \ref{fig:shocks}). There is currently no theory for $\lambda_*$, but prior work suggests $\lambda_*$ increases from about 0.45 to about $0.6$ as $m$ increases from 5 to $\infty$ \citep{Reinelt1985,Rakotomalala1997,Videbaek2020}. Since the shock height is determined by local dynamics, it will become constant as the front moves some $O(1)$ multiple of $\hinf$, which is on a much shorter timescale than the evolution of the whole flow. 

\medskip

To summarise, we have shown in this section that radial intrusions into a Hele-Shaw cell with or without a shock are stable to axisymmetric perturbations, with all perturbations decaying like $\ee^{-\tau}=\hat{t}/t$. We have in (\ref{solSteady}) the shape $X_0(\lambda)$ of a steady axisymmetric similarity solution. We now proceed to the central point of the paper, a linear stability analysis of this base state to determine the growth rate of possible fingering instabilities. 

\section{Linear stability analysis}\label{sec:linearstability}

\subsection{Formulation of the equations}\label{sec:linearstability-eqns}

We wish to consider small non-axisymmetric perturbations to the axisymmetric similarity solution of \S\ref{sec:kwave}. For simplicity, we will assume that any frontal shock for $m>{3\over2}$ is a contact shock and note that this gives the smallest shock height and plausibly the smallest tendency to instability. We return to the case of undercompressive shocks in \S\ref{sec:under}.

Introducing the function $\Phi=\xi q_\xi$ for convenience, we can write (\ref{nondimPDE}) in the form
\begin{subequations}\label{phieqns}
\begin{equation}\tag{\ref{phieqns}{\it a}--{\it b}}
\Phi=-\scM\xi{\partial p\over \partial \xi},\qquad\qquad
\xi{\partial \Phi\over \partial \xi}=\scM{\partial^2 p\over \partial \theta^2}+O(2),
\end{equation}
\begin{equation}\tag{\ref{phieqns}{\it c}}
\qquad{\partial \lambda\over \partial\tau}+{2\scF'\Phi-\xi^2\over 2\xi}{\partial \lambda\over \partial\xi}=O(2)~~\text{for}~~\xi<\xi_*,
\end{equation}
\end{subequations}
where $O(2)$ denotes terms proportional to $(\partial \scM/ \partial \theta)(\partial p/ \partial \theta)$ in (\ref{phieqns}{\it b}) and $q_\theta(\partial \lambda/ \partial \theta)$ in (\ref{phieqns}{\it c}) that are both quadratically small in the perturbation and can thus be neglected in a linear analysis. Neglecting the $O(2)$ term,  (\ref{phieqns}$c$) implies that $\lambda$ is constant to linear order along radial characteristics defined by
\begin{equation}\label{xi2}
	\tdiff{\xi^2}{\tau}+\xi^2=2\scF'\Phi 
	\,,\qquad \tdiff{\theta}{\tau}=0\,.
\end{equation}

The axisymmetric base state is given by $\xi=X_0(\lambda)$, $p=P_0(\lambda)$ and $\Phi=\Phi_0=1$, where
\begin{equation}
	X_0=\left\{2\scF'(\lambda)\right\}^{1/2},\quad
	\scM P_0'=-\scX(\lambda)\quad\text{and}\quad
	\scX(\lambda)\equiv\frac{X_0'}{X_0}=\frac{\scF''}{2\scF'}.
\label{baseState}
\end{equation}
The base state and the functions $\scM$, $\scF$ and $\scX$ are all given as functions of $\lambda$. Hence, it is again convenient to use $\lambda$ as the independent radial variable in place of $\xi$, and to pose the perturbation expansion in $\xi<\xi_*(\theta,\tau)$ in the form
\begin{subequations}\label{expansion}
\begin{align}
	\xi(\lambda,\theta,\tau)&=X_0(\lambda)+X_1(\lambda)\ee^{{\ii}k\theta+\sigma\tau}+\cdots,
	\\
	p(\lambda,\theta,\tau)&=P_0(\lambda)+P_1(\lambda)\ee^{{\ii}k\theta+\sigma\tau}+\cdots,
	\\
	\Phi(\lambda,\theta,\tau)&=\hspace{6.7mm}1+\Phi_1(\lambda)\ee^{{\ii}k\theta+\sigma\tau}+\cdots,
	\end{align}
\end{subequations}
where $\sigma$ is the growth rate. The azimuthal wavenumber $k$ takes integer values for $2\pi$-periodicity, but can be treated as a continuous variable for convenience without loss of generality. We neglect terms that are quadratic or higher in the 
perturbation quantities $X_1$, $P_1$ and $\Phi_1$. 

Applying the chain rule to the transformation from $(\xi,\theta,\tau)$ to $(\lambda,\theta,\tau)$, we transform the derivatives in (\ref{phieqns}{\it a,b}) using
\begin{equation}
\xi{\partial \over \partial \xi}=\frac{\xi}{\partial\xi/\partial\lambda}{\partial \over \partial \lambda}\quad\text{and}\quad
\left({\partial \over \partial \theta}\right)_\xi=\left({\partial \over \partial \theta}\right)_\lambda - \frac{\partial\xi/\partial\theta}{\partial\xi/\partial\lambda}{\partial \over \partial \lambda}\,.
\label{derivs}
\end{equation}
We then substitute the expansion (\ref{expansion}) into (\ref{phieqns}{\it a,b}) and (\ref{xi2}), linearise the result, and use (\ref{baseState}) to simplify the equations further. After some algebra, we obtain 
\begin{equation}
\scX\Phi_1+\Big({X_1\over X_0}\Big)' =-\scM P_1',\qquad\qquad
{\Phi_1'}=-k^2 \scX\Big(\scM P_1+{X_1\over X_0} \Big)
\label{linear}
\end{equation}
and
\begin{equation}
\label{linearX}
2(\sigma+1){X_1\over X_0}=\Phi_1\,.
\end{equation}

The special case $\sigma=-1$ provides stable perturbations, such as the axisymmetric perturbations of \S\ref{sec:kwave}, and it will not be considered further. If $\sigma\neq-1$, we can use (\ref{linearX}) to eliminate $X_1/X_0$ from (\ref{linear}) and obtain the coupled system
\begin{subequations}\label{firstorder}
\begin{equation}\label{ODE}
	\begin{pmatrix}
		\scM P_1' \\ \Phi_1'
	\end{pmatrix}
	=
	\scX
	\begin{pmatrix}
		\frac{k^2}{2(1+\sigma)} & \frac{k^2}{4(1+\sigma)^2}-1 \\
		-k^2 & -\frac{k^2}{2(1+\sigma)}
	\end{pmatrix}
	\dotProd
	\begin{pmatrix}
		\scM P_1 \\ \Phi_1
	\end{pmatrix}.
\end{equation}
As shown in Appendix \ref{app:boundary}, the general boundary conditions (\ref{nondimBCs}) reduce to the  boundary conditions
\begin{equation}\label{odebcs}  
\frac{P_1}{\Phi_1}=\frac{1}{k}-\frac{1}{2(1+\sigma)}
~~\text{at}~~\lambda=\lambda_* \,,\qquad
\Phi_1\to0~~\text{as}~~\lambda\to1 \,
\end{equation}
\end{subequations}
on (\ref{ODE}). Equations (\ref{firstorder}$a,b$) are a linear homogeneous system, which constitutes an eigenvalue problem to determine the growth rates $\sigma(k;m)$ of perturbations with radial structure given by eigenfunctions $P_1$ and $\Phi_1$. It can be solved numerically in this form.

Alternatively, we can eliminate $P_1$ to obtain a second-order equation for $\Phi_1$:
\begin{subequations}\label{secondorder}
\begin{equation} \label{PhiODE}
\scX\scM\left({\Phi'_1\over \scX\scM}\right)'= 
k^2 \scX^2\left( 1+{\scN\over \sigma+1} \right)\Phi_1 \,,
\text{~~where~~}\scN(\lambda)\equiv{\scM'\over 2\scM\scX} \,,
\end{equation}
with boundary conditions
\begin{equation} \label{PhiBCs}
	\frac{\Phi_1'}{k^2\scX_*}=\left(\frac{\scM_*-1}{2(1+\sigma)}-\frac{\scM_*}{k}\right)\Phi_1
	\quad\text{at}\quad\lambda=\lambda_*,\qquad
	\Phi_1\to0 ~\text{as}~~\lambda\to1.
\end{equation}
\end{subequations}
This second-order form is convenient for WKB analysis of the limit $k\to\infty$. 

Though it is slightly unusual for the eigenvalue $\sigma$ to appear in the boundary condition (\ref{PhiBCs}) as well as the differential equation (\ref{PhiODE}), this second-order form is sufficiently close to a standard Sturm--Liouville eigenvalue problem to expect, as proves to be the case (see Appendix \ref{sec:SL}), that there is a discrete spectrum of eigenmodes for each $k$ and $m$. We label these modes by an integer $n\geq0$, which is equal to the number of zeros of the eigenfunction away from the zero boundary condition at $\lambda=1$ or, equivalently, at $\xi=0$ (see figure \ref{fig:pq} for example).

\subsection{Numerical solution and results}\label{sec:linearstability-results}

We solved the boundary-value problem (\ref{firstorder}) numerically using continuation methods implemented with the software package AUTO-07P (freely available at http://indy.cs.concordia.ca/auto/). The strategy for obtaining an eigenmode is analogous to that detailed in \citet{Ribe06}: start with a guess for $\sigma$ and a nonzero solution of (\ref{ODE}) that satisfies one of the homogeneous boundary conditions (\ref{odebcs}) but won't, in general, satisfy the other; then use continuation to slowly impose the other boundary condition, keeping the solution nonzero and allowing $\sigma$ to vary slowly until it reaches an eigenvalue at the point the second boundary condition is satisfied. Having obtained the eigenmode for one set of parameters, continuation can again be used to track its variation with $k$ and $m$. We present results for the first three eigenmodes, $n=0,1,2$, but, given the nature of the spectrum, it is easy to find starting values of $\sigma$ that give the higher eigenmodes. 

\begin{figure}
	\centering
	\includegraphics[width=0.7\textwidth]{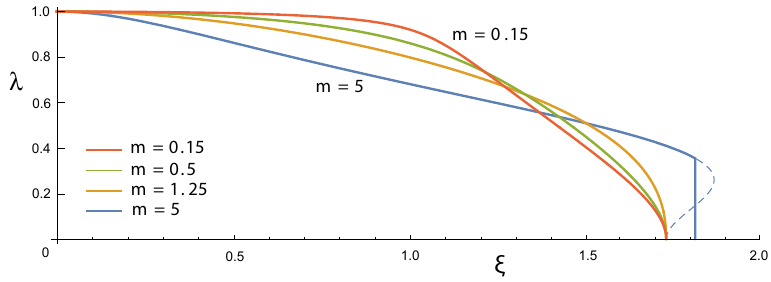}
	\caption{The axisymmetric base-state profiles $X_0(\lambda)$, as given by (\ref{baseState}), of self-similar solutions for intrusions with viscosity ratios $m\in\{0.15,0.5,1.25,5\}$. For $m=5$ there is a frontal shock at $\xi_*=1.815$ of height $\lambda_*=0.354$, rather than the unphysical non-monotonic profile (dashed) that would be predicted  by ignoring the crossing of characteristics.}
	\label{fig:base}
\end{figure}

Figure~\ref{fig:base} shows the analytical base-state profiles (\ref{baseState}) for various values of $m$. For $m<{3\over2}$ the profile has a rounded nose with  the tip position at $\xi_*=\sqrt3$ as determined by a combination of the centreline velocity for $\lambda=0$ and radial spreading. For $m\ll1$ the more viscous intruding fluid is lubricated by the less viscous ambient fluid near the origin and the intrusion there is wider and closer to plug flow than for $m=1$. For $1<m<{3\over 2}$ the lower viscosity intrusion is narrower near the origin, and wider near the rounded nose than for $m=1$. For $m>{3\over2}$ there is a frontal shock, which we are assuming has height given by (\ref{lambdaCrit}). 

In the results below, we will use viscosity ratios $m\in\{0.15,1.25,5\}$ as illustrative examples of the stability behaviours found in the three distinct cases: a more viscous intrusion ($m<1$), a less viscous intrusion without a shock ($1<m<{3\over 2}$) and a less viscous intrusion with a shock ($m>{3\over2}$). We observe briefly that $m=1$ (equal viscosities) is a very special case as the lack of any viscosity differences means that the flow is always radial with flux $\vect{q}=\xi^{-1}\vect{e}_\xi$ and a parabolic (Poiseuille) profile. Hence, there is no perturbation flow ($P_1=\Phi_1=0$), the interface is simply a passive tracer in the base-state flow, and any perturbations to $X_0(\lambda)$ decay purely kinematically with $\sigma=-1$.

\begin{figure}
	\centering
	\includegraphics[width=0.75\textwidth]{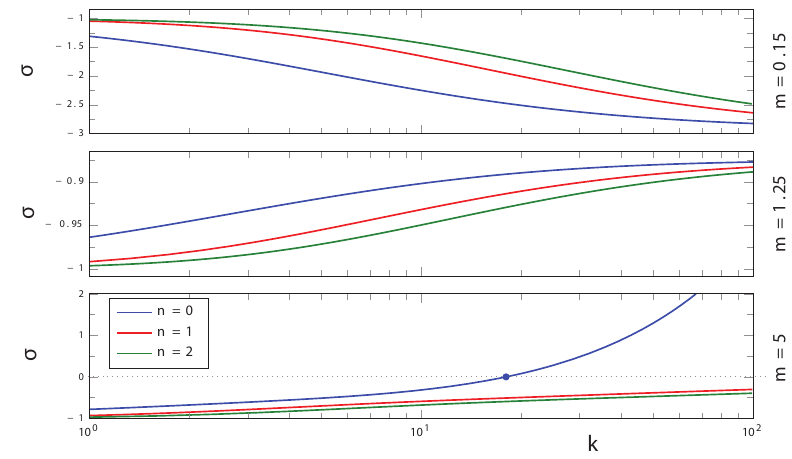}
	\caption{The growth rates $\sigma$ corresponding to the first three eigenmodes $n\in\{0,1,2\}$ with viscosity ratios $m\in\{0.15,1.25,5\}$ as functions of the azimuthal wavenumber $k$. For $m=5$, the fundamental mode $n=0$ is unstable if $k$ exceeds a critical value $\approx 18$ where $\sigma=0$ (blue dot).}
	\label{fig:sigma-k}
\end{figure}

Figure~\ref{fig:sigma-k} shows the growth rates $\sigma$ of the first three eigenmodes $n\in\{0,1,2\}$ for the three illustrative viscosity ratios as functions of the azimuthal wavenumber $k$. Some general observations can be understood physically. First, for all of the eigenmodes, $\sigma\to-1$ as $k\to0$, which reflects the result $\sigma=-1$ for axisymmetric perturbations in \S\ref{sec:kwave}. (Recall, we are treating $k$ as a continuous variable for convenience, rather than imposing $2\pi$-periodicity.) A second, related observation is that in each graph $\sigma$ becomes closer to $-1$ as $n$ increases. Increasing $n$ corresponds to increasing the number of zeros in the eigenfunctions and hence to increasing the amount of radial structure. We can reasonably expect that, as $n\to\infty$ for fixed $k$, the radial structure dominates the azimuthal variation and therefore the growth rate again approaches the $\sigma=-1$ result for axisymmetric perturbations. Third, $\sigma<-1$ for $m<1$ and $\sigma>-1$ for $m>1$. This is consistent with the observation that $\sigma=-1$ for all perturbations when $m=1$, and is also consistent with intuition derived from the Saffman--Taylor instability mechanism that pushing a more viscous fluid into less viscous fluid tends to be stable, whereas pushing a less viscous fluid into a more viscous fluid tends to promote instability. Nevertheless, $m=1$ is definitely stable ($\sigma=-1$) and so $m>1$ is {\em not} sufficient to produce instability, as can been seen, for example, in figure~\ref{fig:sigma-k} for $m=1.25$.

For $m=0.15$, all modes are stable with $\sigma<-1$, the fundamental mode $n=0$ is the most stable, and as $k$ increases the perturbations become more stable.
For $m=1.25$, all modes are again stable, but with $-1<\sigma<0$, the fundamental mode $n=0$ is the least stable and, though the perturbations become less stable as $k$ increases, the growth rates appear to tend to a limit that is still negative as $k\to\infty$. For $m=5$, which has a base state with a shock, modes $n=1,2$ are again stable with $-1<\sigma<0$. The crucial difference for $m=5$ is that the fundamental mode $n=0$ becomes unstable at $k\approx18$ and the growth rate increases rapidly as $k\to\infty$. In \S\ref{sec:largek} we show that the instability mechanism is essentially the Saffman--Taylor mechanism acting on the jump in mobility at the frontal shock.

\begin{figure}
	\centering
	\includegraphics[width=0.85\textwidth]{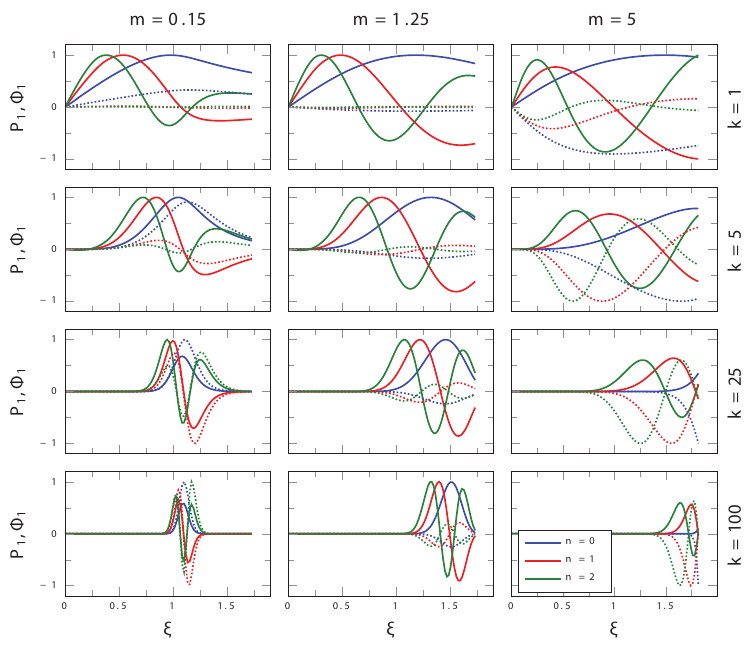}
	\caption{Numerical solutions for the first three eigenmodes $n\in\{0,1,2\}$ in terms of the perturbation pressure $P_1$ (solid) and perturbation flux $\Phi_1$ (dotted) for wavenumbers $k\in\{1,5,25,100\}$ and viscosity ratios $m\in\{0.15,1.25,5\}$. The nose position is at $\xi_*=\sqrt{3}$ for $m\in\{0.15,1.25\}$ and at $\xi_*\approx1.81$ for $m=5$.}
	\label{fig:pq}
\end{figure}

Figure~\ref{fig:pq} shows the radial structure of the first three eigenmodes of the pressure perturbation $P_1$ and flux perturbation $\Phi_1$ as functions of the radial similarity variable $\xi$. The panels show solutions for the three illustrative viscosity ratios and for four azimuthal wavenumbers $k\in\{1,5,25,100\}$. As expected, the number of zeros (additional to $\xi=0$) increases with mode number $n$. We note that $\Phi_1$ is approximately in phase with $P_1$ for $m=0.15$, but has approximately the opposite phase (sign) for $m=1.25$ and $m=5$. From the form of the matrix in (\ref{ODE}), it can be inferred that this is largely a consequence of the sign of the factor $1/(\sigma+1)$, which is different for $m>1$ and $m<1$.

For $k=1$ (sideways displacement, perturbations $\propto\cos\theta$) the figure shows that the eigenmodes giving relaxation back to axisymmetry have a comparable length scale to the full extent of the intrusion. As $k$ increases, the eigenmodes become increasingly localised radially. In Appendix \ref{AppB} we show that as $k\to\infty$ for $m<{3\over2}$ the eigenmodes become localised about an interior position between the origin and the nose; for $m>{3\over2}$ the eigenmodes become localised near the frontal shock.

\begin{figure}
	\centering
	\includegraphics[width=0.7\textwidth]{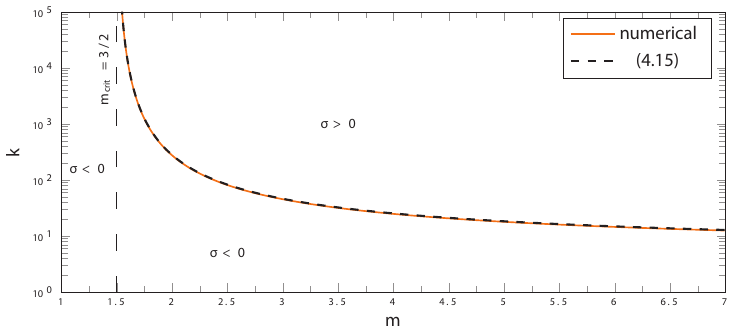}
	\caption{The marginal-stability contour $\sigma=0$ for the fundamental mode $n=0$ in the $(m,k)$-plane compared to the asymptotic result (\ref{bigk-marginal}) for $k\gg1$.  }
	\label{fig:marginal}
\end{figure}

Figure \ref{fig:sigma-k} showed that for $m=5$ the fundamental mode is unstable for $k\gtrsim18$. Figure~\ref{fig:marginal} extends this result by showing the regions in the $(k,m)$-plane where $\sigma>0$ or $\sigma<0$ and the curve of marginal stability where $\sigma=0$.  For $m>{3\over2}$ the fundamental mode is always unstable for sufficiently large $k$, while for $m<{3\over2}$ it is stable for all $k$. In particular, for $1<m<{3\over2}$ the flow is stable to all perturbations, which agrees with experimental observations of stability in this regime, but contrasts with instability in the classical Saffman--Taylor problem for $m>1$. 

As $m$ decreases towards ${3\over2}$ these calculations show that the flow is only unstable to very large $k$, for example $k>10^3$ for $m<1.75$. However, if $k$ is too large ($k\gg r_*/\hinf$) then the horizontal lengthscale $r_*/k$ of perturbations near the front is less than the thickness $2\hinf$ of the Hele-Shaw cell and the horizontal viscous stresses, which are neglected in the lubrication model (\ref{lubrication}), will stabilise the flow and provide a large wavenumber cutoff \citep[cf.][]{Paterson1985}. It follows that as $m$ decreases toward ${3\over2}$ the instability would only be manifest at very large $r_*/\hinf$ and would thus be difficult to observe experimentally in practice.

Figure~\ref{fig:marginal} also shows asymptotic results for the large-wavenumber limit $k\to\infty$, which are derived in the next subsection. The excellent agreement with the numerical results supports the accuracy of the calculations.

\subsection{Analysis of the limit $k\gg1$ for $m>{3\over 2}$ and $n=0$}\label{sec:largek}

Numerically, instability occurs only for $n=0$, sufficiently large $k$ and $m>{3\over2}$ (figures \ref{fig:sigma-k} and \ref{fig:marginal}). We now pursue an analysis of the limit $k\to\infty$ to confirm these results and examine the nature of the instability as $m\to{3\over2}$. Numerically, it appears that $\sigma\sim k$ as $k\to\infty$ and we note this is also true for the Saffman--Taylor instability in the limit of vanishing surface tension. 

For convenience of notation, we define
\begin{equation}\label{Tdef}
T={k\over 2(1+\sigma)}\,,
\end{equation}
which we anticipate will be an $O(1)$ quantity if $\sigma\sim k$. Equations (\ref{secondorder}{\it a,b}) can then be written in the form
\begin{subequations}\label{bigk}
\begin{equation} \label{bigkODE}
\Phi_1''- k^2  \scX_0^2\Phi_1 =
\Big({\scX_0'\over \scX_0}+{\scM'\over \scM}\Big)\Phi_1'
 +k\scX_0 {T\scM'\over \scM} \Phi_1 \,,
\end{equation}
\begin{equation} \label{bigkBC}
	\frac{\Phi_1'}{k\scX_*}=\left(T\scM_*-T-\scM_*\right)\Phi_1
	\quad\text{at}\quad\lambda=\lambda_*,\qquad
	\Phi_1\to0 ~\text{as}~~\lambda\to1.
\end{equation}
\end{subequations}
We make the usual WKB assumption that $\Phi_1=\exp[S(\lambda)]$ with $S=kS_0+S_1+O(k^{-1})$ and substitute into (\ref{bigkODE}) to obtain
\begin{equation} \label{phase}
k^2S_0'{}^2- k^2  \scX_0^2+ kS_0''+2kS_0'S_1'=kS_0'\Big({\scX_0'\over \scX_0}+{\scM'\over \scM}\Big)  + k\scX_0 {T\scM'\over \scM} +O(1)\,.
\end{equation}

At $O(k^2)$ we have the decaying solution $S_0'=+\scX_0$. (Note $\scX_0<0$ so this solution has $\Phi_1\to0$ as $\lambda\to1$.) This also implies that $kS_0''=kS_0'\scX_0'/\scX_0$ in (\ref{phase}). At $O(k)$ the remaining terms simplify to 
\begin{equation}
2S_1'= (1+T) {\scM'\over \scM} \quad \Rightarrow\quad S_1= {1+T\over 2} \ln\scM +c\,.
\end{equation}
We note that we can substitute the solutions for $S_0$ and $S_1$ into the WKB ansatz to obtain a uniformly asymptotic expression $\Phi_1=A\scM^{(1+T)/2} (X_0/X_{0*})^k$ for the structure of the fundamental eigenmode, which includes the effect of radial mobility variations. 

The boundary condition (\ref{bigkBC}) at $\lambda=\lambda_*$ yields
\begin{equation}
\frac{S_0'}{\scX_*}+{S_1'\over k\scX_*}+O(k^{-2})=T\scM_*-T-\scM_* \,,
\end{equation}
where $S_0'$ and $S_1'$ are now known. This equation can be rearranged using (\ref{Tdef}) to give the desired asymptotic result for the growth rate
\begin{equation}\label{bigkn0}
\sigma(k;m)\sim{k\over 2}{\scM_*-1\over \scM_*+1}-1+ {1\over 2|\scX_*|}{\scM'_*\over (\scM_*+1)^2}+O(k^{-1}) \,.
\end{equation}

\begin{figure}
	\centering
	\includegraphics[width=0.7\textwidth]{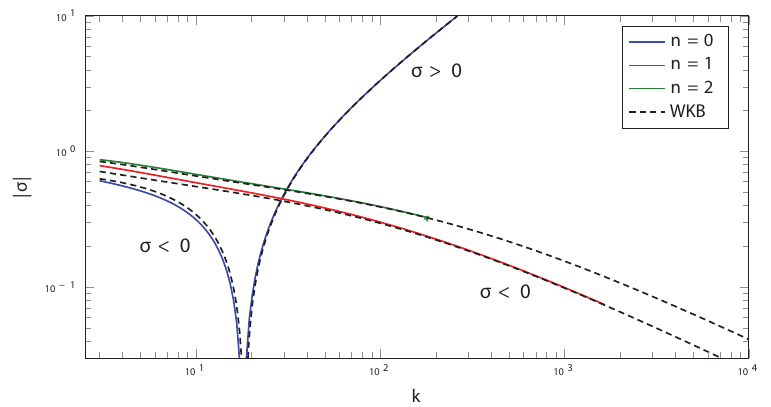}
	\caption{Comparison of the numerically computed growth rate $\sigma(k)$ for $n\in\{0,1,2\}$ and $m=5$ (solid colour) with the corresponding WKB solution (dashed). Note that $\sigma$ changes sign for $n=0$ at $k\approx18$.}
	\label{fig:largek-sigma}
\end{figure}

As shown in figure \ref{fig:largek-sigma}, (\ref{bigkn0}) gives a very good approximation to the full numerical result, even at moderate $k$. The first term agrees with a simple analysis of Saffman--Taylor instability at the front of an intrusion of uniform thickness and hence uniform mobility $\scM_*>1$ \citep[see, e.g.,][] {Saffman1958,Lajeunesse2001,Videbaek2020}. The second term, $-1$, reflects the stabilising effect of radial geometry \citep[cf.][]{Wilson1975,Paterson1981,Paterson1985}. 

The third term, involving $\scM_*'$, is new and describes the effects of the base-state variation of the intrusion thickness away from the front. The term is positive as the mobility $\scM$ is an increasing function of $\lambda$ for $m>1$ and so $\scM'_*>0$. Hence it represents an additional destabilisation relative to the Saffman--Taylor result for an intrusion of uniform thickness in radial geometry. This may be understood physically as the effect of slightly greater mobility just behind the frontal shock, which facilitates the growth of perturbations.
The size of the third term varies only slowly with $m$ from $1\over 4$ at $m={3\over 2}$ to $3\over{16}$ as $m\to\infty$.  

By setting $\sigma=0$ in (\ref{bigkn0}), we can also obtain an asymptotic estimate for the curve of marginal stability. We substitute for $\scM_*$ and $\scX_*$ from (\ref{MFdef}) and (\ref{baseState}), and rearrange (\ref{bigkn0})
to obtain 
\begin{align}\label{bigk-marginal}
	k(m)\approx\frac{3}{(m-1)\lambda_*^3}+\frac{3+2(m-1)\lambda_*^3}{2+(m-1)\lambda_*^3}\,.
\end{align}
This result is asymptotic as $m\to{3\over2}_+$ since this gives $\lambda_*\to0$ and $k\to\infty$, and it agrees remarkably well with the numerically calculated curve of marginal stability (figure~\ref{fig:marginal}), even for relatively small wavenumber $k$.

The WKB analysis for $n=0$ with $m>{3\over2}$ has thus provided excellent confirmation of the numerical results and gives asymptotic expressions as $k\to\infty$ for the unstable eigenmode, growth rate and marginal stability curve that include the leading-order effects of the radial variation in the mobility $\scM$ of the base state. In Appendix \ref{AppB} we analyse the limit $k\to\infty$ for $n>0$ with $m>{3\over 2}$ and for $n\geq0$ with $m<{3\over 2}$, which again provides good confirmation of the numerical results. For $1<m<{3\over2}$ we find that $-1<\sigma<-{3\over4}$.

\subsection{Undercompressive shocks}\label{sec:under}

A formal stability analysis of undercompressive shocks is somewhat complicated by the need to split the intrusion into a central region where $\lambda>\lambda_*$ that is reached by characteristics from the origin and an annular region where $\lambda=\lambda_*$ that lies between the shock and the central region. The equations in the annular region are similar to those ahead of the shock, but with mobility $\scM_*$. As noted earlier, there is currently no theory for $\lambda_*$. Fortunately, we can, nevertheless, derive a simple result from the large-$k$ analysis in the previous section. 

Recalling that $\scX=X_0'/X_0$, we can rewrite $\scM'/\scX$ in (\ref{bigkn0}) as $X_0\,{\rm d}\scM/{\rm d}X_0$. For an undercompressive shock this derivative is zero in the annular region of constant $\lambda$ and thus
\begin{equation}\label{bigkunder}
\sigma(k;m)\sim{k\over 2}{\scM_*-1\over \scM_*+1}-1 \,.
\end{equation}
This result is the same as for an entirely uniform-thickness intrusion \citep{Paterson1985,Videbaek2020}. It appears here as the asymptotic result, despite the non-uniform central region, because the asymptotic radial eigenfunction $\Phi_0\propto (X_0/X_{0*})^k$ is concentrated near the front, and lies in the uniform annular region (cf. the final panel of figure \ref{fig:pq}).  

The loss of the third term from (\ref{bigkn0}) makes the disturbance more stable, but the fact that $\lambda_*>\lambda_c$ for an undercompressive shock makes the disturbance less stable. The net effect depends on $\lambda_*$. However, provided $\lambda_*\to0$ as $m\to{3\over2}$, the marginal stability estimate from (\ref{bigkunder}) also has $k\to\infty$.

\section{Discussion and conclusions}\label{discussion}

Using lubrication theory, we have derived the equations governing the shape of a fluid tongue intruding from a point source into a Hele-Shaw cell filled with another fluid of the same density but differing viscosity, neglecting both diffusion and surface tension. For the case of perfectly axisymmetric flow, changing variables to the relative fluid fraction $\lambda$ allowed an explicit solution from initial conditions, from which we could show that the initial-value problem approaches a late-time similarity solution with all relative perturbations decaying like $t^{-1}$. This similarity solution is the axisymmetric equivalent of that found by \citet{Yang1997} for rectilinear flow, and it grows like $t^{1/2}$ instead of $t$. As in \citet{Yang1997}, if the viscosity ratio $m$ exceeds ${3\over2}$  there must be a frontal shock of height $\lambda_*\geq\lambda_c$.

Our linear stability analysis of non-axisymmetric perturbations to this self-similar base state has provided clear theoretical confirmation of the previously stated hypothesis that the frontal shock is crucial for the development of instability in miscible Hele-Shaw displacement with negligible diffusion: axisymmetric flows with a viscosity ratio $m<{3\over2}$, and hence without a shock, were shown to be stable; axisymmetric flows with a viscosity ratio $m>{3\over2}$, and hence with a shock, were shown to be unstable. In particular, this means that intrusions of less viscous fluid into a more viscous ambient can be stable, provided the viscosity ratio does not exceed ${3\over2}$. 

The analysis gives an indication why the critical viscosity ratio is $m={3\over2}$ and not $m=1$ as for immiscible displacement. For miscible displacement with $m=1$ the fluid interface is just a passive tracer, which is advected by the base-state radial Poiseuille flow. The kinematics are the same as described in \S\ref{sec:kwave} and give stable behaviour with $\sigma=-1$. Crucially, the radial gradient of the mobility $\scM$ remains finite for $1\leq m<{3\over2}$ and so $\sigma$ varies continuously from $\sigma=-1$ as $m$ varies from 1. As might be expected physically, the flow does becomes less stable as $m$ increases but, as shown in Appendix \ref{AppB}, $\sigma$ actually remains less than $-{3\over4}$ for any $m<{3\over2}$ and hence the flow remains stable. (By contrast, immiscible displacement has a sharp jump in mobility at the front for $m>1$, and this causes a significant change in the behaviour of $\sigma$ at large $k$, which drives the Saffman--Taylor instability.)

For $m>{3\over2}$ there must be a frontal shock and a mobility jump, and we find the expected Saffman--Taylor-like instability. Our analysis puts this expectation on a firm theoretical footing by including the effects of the radial variation of the base state. There is a discrete spectrum of radial eigenmodes with differing numbers of zeros, and only the fundamental mode $n=0$ can be unstable. It is unstable for sufficiently large azimuthal wavenumber $k$ and, within the lubrication model, is most unstable as $k\to\infty$.  This divergence for very large $k$ could be regularised \citep[see, e.g.,][]{Paterson1985,Dias2013,Nagel2013} by re-including neglected horizontal stresses, which would stabilise wavelengths less than the cell thickness and predict the most unstable wavelength to be a few times the cell thickness.

Comparison of wavelengths to the cell thickness is also essential for  prediction of the onset radius at which the instability becomes manifest in experiments, particularly as $m$ decreases towards ${3\over2}$. As shown in figure \ref{fig:marginal}, the sufficiently large $k$ for instability is predicted to increase rapidly as $m$ decreases and it diverges as $m\to{3\over2}$. For example, $k>10^3$ is necessary for instability with $m=1.75$. For the unstable wavelengths to exceed the cell thickness, one also needs $r_*>\hinf k$ making the instability more difficult to observe in a given experimental cell. The predicted divergence in the onset radius is consistent with experimental data in \citet{Videbaek2020} and with the difficulty in observing instability for $m$ less than about 2--3 \citep{Lajeunesse1997,Bischofberger2014}.  (A further contributing factor may be the time take to form the frontal shock structure: for $m=2$ the velocity of even the fastest characteristic exceeds the centreline velocity by less than 1\%.)

In addition to the numerical results, we also found asymptotic solutions to the linear stability problem by using a WKB analysis for large $k$. These confirm the numerical results and provide useful analytic expressions for the growth rates, for example, the remarkably good marginal stability curve (\ref{bigk-marginal}) for a contact shock or the asymptotic growth rate (\ref{bigkunder}) of an undercompressive shock. 

It would be nice to have a better understanding which of the two theoretical shock structures applies for $m<5$, and why. We note, however, both from above and from \citet{Goyal2006}, that for $m<2-3$ the front is likely very slow to become quasisteady (even in a tube where it is stable), and that very large, or infinite, P\'eclet number may be required for diffusion to remain negligible. Full direct numerical simulations or experiments could be challenging. The effects of diffusion (physical or numerical) increase with radial distance in an axisymmetric geometry since the velocity $\dot r_*$ of the front decreases like $t^{-1/2}$. \citet{Goyal2006} studied diffusive effects on instability numerically in a linear geometry and found that the most unstable wavelength and the growth rate increase only slightly, by of order 10\%, for $m>7$ as the P\'eclet number $2\hinf \dot x_*/D$ increased from 500 to 2000. Similarly, \citet{Videbaek2019} found similarly in radial (and linear) experiments that the most unstable wavelength and also the onset radius showed no discernible trend for $m=5$ as the P\'eclet number at onset ranged from 1000 to 20000, but also that a quite different fingering instability is seen at lower P\'eclet numbers. The boundary between their low- and high- P\'eclet number regimes increases as $m$ decreases, which might be linked to the increase in the marginally stable wavenumber, and hence onset radius, shown here (figure \ref{fig:marginal}) as $m$ decreases towards the critical value $m={3\over2}$. While questions remain over the detailed effects of diffusion, these studies are at least indicative of a high-Peclet-number regime in which diffusion can be neglected. Perhaps the shock structure could be investigated by Stokes-flow calculations in this limit.

We have concentrated here on the case of radial flow in part because of its experimental relevance and in part because the radial geometry allows fully separable eigenfunctions with a certain radial structure in similarity space, fixed azimuthal wavenumber and power-law time-dependence. (Analysis of unidirectional flow is significantly more difficult.) To summarise our main conclusions: the case $m = 1$ is kinematically stable with algebraic decay like $t^{-1}$; intrusions with $1 < m < {3\over2}$ are stable, though less so; intrusions with $m > {3\over2}$ are unstable for sufficiently large wavenumber due to the jump in mobility, and hence pressure gradient, at the frontal shock; the instability is hard to observe for $m$ only a little larger than ${3\over2}$.


\bigskip

{\bf Funding:}
T.-F.D. was supported by an Engineering and Physical Sciences Research
Council PhD studentship. 

{\bf Data availability statement:}
The research data supporting the findings of this study 
are available within the article.

{\bf Competing interests:}
The authors declare none.

{\bf Author ORCID:} John R. Lister, https://orcid.org/0000-0002-8978-2672

%
\appendix
\section{Boundary conditions on the perturbations}\label{app:boundary}

The perturbation equations (\ref{ODE}) are a  second-order system of linear ordinary differential equations and hence we expect two boundary conditions. Physically, these are given by a matching condition to the flux and pressure distribution ahead of the nose and a condition on there being no perturbation flux at the origin. 
Mathematically, we solve the system ahead of the nose analytically to find a matching condition via (\ref{nondimBCs}$b$--$d$) and we use a local expansion near the origin to find an appropriate boundary condition from (\ref{eqn:BC_influx}). 

Ahead of the nose ($\xi>\xi_*$), we have $\lambda\equiv0$ and must pose a perturbation expansion of the form 
\begin{subequations}\label{xiexpand}\begin{align}
	p(\xi,\theta,\tau)&=p_0(\xi)+p_1(\xi)\ee^{\ii k\theta+\sigma\tau}+\cdots,
	\\
	\Phi(\xi,\theta,\tau)&=1+\phi_1(\xi)\ee^{\ii k\theta+\sigma\tau}+\cdots,
\end{align}\end{subequations}
rather than (\ref{expansion}). 
Since $\lambda=0$, we have $\scM=1$, $\vect{q}=-\grad p$ and $\nabla^2 p=0$.
The base state  $\Phi=1$ gives $p_0=-\ln\xi$ to within an unimportant additive constant.  
Since the perturbation term in (\ref{xiexpand}$a$) also satisfies Laplace's equation, we must have $p_1\propto\xi^{\pm k}$. Assuming $k>0$ for definiteness, the condition (\ref{eqn:BC_infty}) of decaying flow in the far-field rules out $\xi^{+k}$, and hence we obtain
\begin{equation}
\label{p1phi1}
p_1(\xi)=\amp \xi^{-k}\quad\text{and}\quad \phi_1(\xi)=\amp k\xi^{-k} \,,
\end{equation}
where $\amp$ is the (small) perturbation amplitude ahead of the nose.

Since $q_\theta n_\theta$ is quadratically small in the perturbation, the frontal boundary conditions (\ref{nondimBCs}$c$) and (\ref{nondimBCs}$d$) reduce to $\Phi$ and $p$ being continuous across $\xi=\xi_*$. As discussed in \S\ref{sec:shocks}, contact shocks tend to their equilibrium height like $\ee^{-\tau}$. Hence for $\sigma\ne-1$ we can assume that the frontal height $\lambda_*$ is not perturbed, and so the frontal position from the expansion (\ref{expansion}{\it a}) is $\xi_*=X_0(\lambda_*)+X_1(\lambda_*)\ee^{\ii k\theta+\sigma\tau}$. We substitute this position into (\ref{xiexpand}) and equate the linearised results to the expansions (\ref{expansion}{\it b,c}) to obtain
\begin{equation}
\label{cont}
P_1=p_1(X_0)+X_1{{\rm d}p_0\over {\rm d}\xi}=p_1(X_0)- {X_1\over X_0} 
\quad\text{and}\quad
\Phi_1=\phi_1(X_0) 
\quad\text{at}\quad \lambda=\lambda_*\,.
\end{equation}
We use (\ref{linearX}) to replace $X_1/X_0$ by $\Phi_1/2(1+\sigma)$. We then divide 
(\ref{cont}$a$) by (\ref{cont}$b$) to obtain the desired matching condition 
\begin{equation}\label{eqn:BC-cty}
	\frac{P_1}{\Phi_1}=\frac{1}{k}-\frac{1}{2(1+\sigma)}
	\quad\text{at}\quad\lambda=\lambda_*,
\end{equation}
which ensures that the solution to (\ref{ODE}) in $x<x_*$ can be matched to the decaying solution (\ref{p1phi1}) in $x>x_*$.

At the origin $\xi=0$, we have $\lambda=1$. Expanding (\ref{solSteady}) about this point yields $\lambda\sim1-{1\over 6}m\xi^2$ as $\xi\to0$ and thus $\scM=m+O(\xi^2)$. We can thus approximate (\ref{phieqns}) by
\begin{equation}\label{origin}
\Phi\sim-m\xi{\partial p\over \partial \xi}\quad\text{and}\quad
\xi{\partial \Phi\over \partial \xi}\sim m{\partial^2 p\over \partial \theta^2}\quad\text{as}\quad\xi\to 0.
\end{equation}
Hence $p$ again satisfies Laplace's equation at leading order and substitution of the expansion (\ref{xiexpand}) again leads to $p_1\propto \xi^{\pm k}$. This time it is the singular solution $\xi^{-k}$ that is ruled out by the origin condition (\ref{eqn:BC_influx}), and hence we obtain 
\begin{equation}
\label{p1phi1_orig}
p_1(\xi)\sim B \xi^{k}\quad\text{and}\quad \phi_1(\xi)\sim -B m k\xi^{k} 
\quad\text{as}\quad\xi\to 0,
\end{equation}
where $B$ is an amplitude. 
Again by equating the expansions (\ref{expansion}) and   (\ref{xiexpand}), this time as $\xi\to0$ or $\lambda\to1$, we obtain
\begin{equation}
\label{orig}
P_1=p_1(X_0)- {X_1\over m X_0} 
\quad\text{and}\quad
\Phi_1=\phi_1(X_0) 
\quad\text{as}\quad \lambda\to1.
\end{equation}
The condition $\Phi_1(1)=0$ is the simplest way of imposing the boundary condition (\ref{eqn:BC_influx}) on (\ref{PhiODE}). Alternatively, 
we again eliminate $X_1/X_0$ and divide the equations to obtain the equivalent condition
\begin{align}\label{BC-origin}
	\lim_{\lambda\to1}\frac{P_1}{\Phi_1}=-\frac{1}{mk}-\frac{1}{2m(1+\sigma)}\,,
\end{align}
which is more convenient numerically for (\ref{ODE}).
Either form of boundary condition ensures regularity of the perturbation as $\xi\to0$.

\section{Analysis of the limit $k\to\infty$ for the stable modes}\label{AppB}

We wish to solve (\ref{secondorder}) as $k\to\infty$ for the cases of $n\geq1$ with $m>{3\over 2}$ and of $n\geq0$ with $m<{3\over 2}$. From figure \ref{fig:sigma-k} we expect that $\sigma\to0$ in the first case and that $\sigma$ tends to a finite negative limit (distinct from $-1$) in the second.

From the form of (\ref{PhiODE}) we expect that $\Phi_1(\lambda)$ varies rapidly on a scale that is $O\big((k\scX)^{-1}\big)$ except perhaps where $\sigma\approx -\scN-1$. On the other hand, the functions $\scX(\lambda)$, $\scM(\lambda)$ and $\scN(\lambda)$ vary relatively slowly over the $O(1)$ interval $\lambda_*<\lambda<1$. Hence we expect $\Phi''\gg \Phi'\scX'/\scX$, $\Phi''\gg\Phi'\scM'/\scM$ and we asymptotically approximate (\ref{secondorder}) by 
\begin{subequations}\label{appsystem}
\begin{equation} \label{appODE}
\Phi_1''= 
k^2 \scX^2\left( 1+{\scN\over \sigma+1} \right)\Phi_1 \,,
\text{~~where~~}\scN(\lambda)={\scM'\over 2\scM\scX} 
= {\scM'\scF'\over \scM\scF''}\,,
\end{equation}
\begin{equation} \label{appBCs}
	\Phi_1=0
	\quad\text{at}\quad\lambda=\lambda_*,\qquad
	\Phi_1\to0 ~\text{as}~~\lambda\to1\,.
\end{equation}
\end{subequations}
For $m>{3\over2}$ the asymptotic approximation of (\ref{PhiBCs}) by (\ref{appBCs}) as $k\to\infty$ follows from  $\scM_*\ne1$, $\sigma=O(1)$ and $\Phi_1'/k^2=O(k^{-1})$; for $m<{3\over2}$ the approximation will be justified 
in \S\ref{AppB2}

To analyse (\ref{appsystem}), we again make the WKB assumption that $\Phi_1=\exp[kS_0+S_1+O(k^{-1})]$ and find straightforwardly that 
\begin{equation} \label{appS0}
S_0'= \pm|\scX|\left( 1+{\scN\over \sigma+1} \right)^{1/2} \,.
\end{equation}
If $1+\scN/(\sigma+1)>0$ everywhere then $S_0'$ is real, the WKB solutions are exponential in character and it is impossible to satisfy the boundary conditions (\ref{appBCs}). Hence, to satisfy the boundary conditions, there must be a region where $1+\scN/(\sigma+1)<0$ and the WKB solutions are oscillatory. For $n=O(1)$ this region must be relatively small since the oscillations are rapid. 

The full solutions to (\ref{appsystem}) are constructed by matching the oscillatory solution in the region where $1+\scN/(\sigma+1)<0$ through the `turning point(s)' where $1+\scN/(\sigma+1)=0$ to exponentially decaying behaviour in the region(s) where $1+\scN/(\sigma+1)>0$. The value of $\sigma$ is determined by the criterion that matching through the turning point results in only the decaying, and not the growing, solution. There are two cases to consider.

\subsection{Case $m>{3\over2}$ and $n\geq1$}\label{AppB1}

For $m>{3\over2}$ it can be shown that $\scN(\lambda)$ varies monotonically from $\scN(1)=0$ to $\scN(\lambda_*)=-1$. 
Hence if $\sigma>0$ or $\sigma<-1$ there is only exponential behaviour and no solution. 
We deduce that $-1<\sigma<0$. Moreover, if the region of oscillatory behaviour is relatively small for $n=O(1)$ then it must be near $\lambda=\lambda_*$ and we must have $|\sigma|\ll1$. 

We expand locally using $|\sigma|\ll1$ and $\scN_*=-1$ to obtain 
\begin{equation} \label{airyarg}
 1+{\scN(\lambda)\over 1+\sigma} = 1+\scN_*(1-\sigma+\dots)+(\scN-\scN_*)(1-\sigma+\dots)=
 \sigma+(\lambda-\lambda_*)\scN'_*+\dots \,,
\end{equation}
where $\scN'_*>0$. At leading order, (\ref{appODE}) reduces to a shifted form of Airy's equation 
\begin{equation} \label{airyeq}
\Phi_1''= 
k^2 \scX^2\big\{ \sigma+(\lambda-\lambda_*)\scN'_* \big\}\Phi_1 \,.
\end{equation}
The condition of matching to a decaying solution as $\lambda\to1$ requires
\begin{equation}\label{airyfn}
\Phi_1\sim {\rm Ai}\Big( \big(k^2\scX_*^2\scN'_*\big)^{1/3}\big\{\lambda-\lambda_*+{\sigma\over \scN'_*}\big\}\Big)\,,
\end{equation}
where Ai($z$) denotes the Airy function.
We note that $\Phi_1'/k^2=O(k^{-4/3})$, which is consistent with our previous approximation of (\ref{PhiBCs}) by (\ref{appBCs}). 

The condition $\Phi_1(\lambda_*)=0$ requires $(k|\scX_*|/\scN'_*)^{2/3}\sigma$ to be one of the roots $z_n$ of ${\rm Ai}(z)=0$.
These roots satisfy ${2\over3}(-z_n)^{3/2}\sim(n-{1\over4})\pi$ as $n\to\infty$ \citep{Abramowitz}, and the formula gives even the first root correct to within 1\%. Using this approximation, we obtain the growth rate of the $n$th eigenmode as
\begin{equation}\label{airysig}
\sigma\sim-\left(\frac{3}{2}\big\{(n-\tfrac{1}{4})\pi\big\} \frac{\scN'_*}{|\scX_*|}\right)^{2/3}k^{-2/3}
	\quad\text{for }n\geq1
\end{equation}
which, as anticipated, is negative and tends to zero as $k\to\infty$.  The local expansion can be continued to calculate an $O(k^{-4/3})$ correction \citep{Dauck}. A global WKB approximation can be obtained by integration of the equations for $S_0$ and $S_1$ away from the turning point and the result of this approximation gives excellent agreement with the full numerical results for $m=5$ and $n=1,2$ shown in figure \ref{fig:largek-sigma}.

\subsection{Case $m<{3\over2}$ and $n\geq1$}\label{AppB2}

For the case $1<m<{3\over 2}$ it can be shown that $\scN(0)=\scN(1)=0$, $\scN(\lambda)<0$ for $\lambda\in(0,1)$ and $\scN(\lambda)$ has a unique minimum $\scN_m=\scN(\lambda_m)$, where $\scN_m\in(-{1\over4},0)$ and depends on $m$.
Hence if $\sigma+1>-\scN_m$ or $\sigma+1<0$ then $1+\scN/(\sigma+1)>0$ everywhere, there is only exponential behaviour and no solution. We deduce that $-1<\sigma<-1-\scN_m$, that the small region of oscillatory behaviour for $n=O(1)$ must be near $\lambda_m$ and that $\sigma+1$ must be just a little smaller than $-\scN_m$.

We expand (\ref{appODE}) near $\lambda_m$ to obtain
\begin{equation}\label{qho}
\Phi_1''
=k^2  \scX_m^2\Big(
{\sigma+1+\scN_m+{1\over2}\scN_m''(\lambda-\lambda_m)^2\over -\scN_m} \Big)\Phi_1 \,, 
\end{equation}
where $\scN''_m>0$ and we have used $\sigma+1\sim -\scN_m$ in the denominator. Equation (\ref{qho}) has the same form as the equation of a quantum harmonic oscillator and the eigenvalues and eigenfunctions are determined similarly by the condition of matching to decaying solutions outside the oscillatory region (i.e.~for $|\lambda-\lambda_m|\gg k^{-1/2}$). On rescaling the standard results for the harmonic oscillator, we obtain the asymptotic growth rate of the $n$th eigenmode as
\begin{equation}\label{qhosig}
\sigma\sim -1-\scN_m+{(2n+1)\big(-\scN_m''/2\scN_m\big)^{1/2}\scN_m\over |\scX_m|} k^{-1}\quad\text{for }n\geq0\,
\end{equation}
which, as anticipated, is just a little smaller than $-1-\scN_m$. The value of $\scN_m$ varies monotonically in $1<m<{3\over 2}$ with $\scN_m\to0$ as $m\to1_+$ and $\scN_m\to-{1\over4}$ as $m\to{3\over2}{}_-$. Thus as $m$ approaches 1 the full range of decay rates $-1<\sigma(k)<-1-\scN_m$ is confined increasingly closely to the $t^{-1}$ behaviour obtained at $m=1$ when the interface is just a passive tracer. As $m$ approaches ${3\over2}$ the range of decay rates expands to rates between $t^{-3/4}$ and $t^{-1}$ at large and small $k$ respectively, but remains well short of instability. (There is a non-uniformity in the double limit $m\to{3\over2}$ as $k\to\infty$, which in principle should allow matching to the behaviours found for $m>{3\over2}$ in \S\ref{sec:largek} and \S\ref{AppB1}.)


For the case $m<1$ we find that $\scN(0)=\scN(1)=0$ again, but $\scN(\lambda)>0$ for $\lambda\in(0,1)$ and $\scN(\lambda)$ has a unique maximum $\scN_m$.
Hence if $\sigma+1<-\scN_m$ or $\sigma+1>0$ then $1+\scN/(\sigma+1)>0$ everywhere and we deduce that $-1-\scN_m<\sigma<-1$. The signs of $\scN_m$, $\scN_m''$ and $\sigma+1$ are the opposite of those for $1<m<{3\over 2}$, but the same argument can be followed through and leads to the same expression (\ref{qhosig}) for the asymptotic behaviour of the growth rates.

The asymptotic analysis again agrees well with the numerical calculations. For example, as $k\to\infty$, $\sigma\to-1-\scN_m$ and the eigenfunctions concentrate around $\xi_m=X_0(\lambda_m)$. For $m=0.15$ this gives $\sigma\to-2.93$ and $\xi_m=1.10$, while for $m=1.25$ this gives $\sigma\to-0.87$ and $\xi_m=1.53$ cf.~figures \ref{fig:sigma-k} and \ref{fig:pq}.

\section{Connection with Sturm--Liouville problems}\label{sec:SL}

To establish a closer connection between the linear-stability problem of \S\ref{sec:linearstability-eqns} and Sturm--Liouville theory, it is convenient to change variables using $\xi=\big\{\scF(\lambda)\big\}^{1/2}$ and $\scX^{-1}{\rm d}/{\rm d}\lambda= \xi\, {\rm d}/{\rm d}\xi$ to rewrite (\ref{secondorder}) as
\begin{equation} \label{xiODE} 
\frac{{\rm d}}{{\rm d}\xi}\left(\frac{\xi}{\scM}\frac{{\rm d}\Phi_1}{{\rm d}\xi}\right)= 
\frac{k^2}{\xi\scM} \left( 1+\frac{\scN}{\sigma+1} \right)\Phi_1 \,,
\text{~~where~~}\scN=\frac{\xi}{2\scM}\frac{{\rm d}\scM}{{\rm d}\xi} \,.
\end{equation}
where $\scM$ and $\scN$ are now implicit functions of $\xi$.
We compare (\ref{xiODE}) with the standard form $(py')'-qy+\lambda w y=0$ and set
\begin{equation}
p(\xi)={\xi\over\scM},\quad q(\xi)=\frac{k^2}{\xi\scM},\quad w(\xi)=\frac{k^2}{ \scM^2}\left|\frac{{\rm d}\scM}{{\rm d}\xi}\right|,\quad \lambda=\frac{{\rm sgn}(m{-}1)}{ 2(\sigma+1)}~\,.
\end{equation}
Here $p$, $q$ and $w$ are continuous on $(0,\xi_*)$, $p$ and $w$ are positive on $(0,\xi_*)$ and the factor ${\rm sgn}(m{-}1)$ accounts for ${\rm d}\scM/ {\rm d}\xi<0$ if $m>1$ and $>0$ if $m<1$. 

For $m<{3\over2}$ we have $\lambda_*=0$. Hence $\scM_*=1$ and the boundary conditions (\ref{PhiBCs}) reduce to 
\begin{equation} \label{SLBCs}
	\Phi_1\to0 ~\text{as}~~\xi\to0,\qquad
\xi\frac{{\rm d}\Phi_1}{{\rm d} \xi}+k\Phi_1=0
	\quad\text{at}\quad\xi=\xi_*	\,.
\end{equation}
Equation (\ref{xiODE}) has a singular point at $\xi=0$, where $p=0$, but it is regular and non-oscillatory, with local solutions $\Phi_1=\xi^{\pm k}$. (The situation is the same as for Bessel's equation.) Thus (\ref{xiODE}) and (\ref{SLBCs}) form a Sturm--Liouville problem and we can conclude that there is a discrete spectrum of distinct eigenvalues $\lambda_n$ with $\lambda_n\to\infty$ as $n\to\infty$ (see, e.g., \citet{Morse1953}, pp.719ff.). It follows that, for fixed $k$, $\sigma_n\to-1$ from above (below) if $m>1$ ($m<1$) as $n\to\infty$, which significantly extends the trends with $n$ seen in figure \ref{fig:sigma-k}. 

For $m>{3\over2}$ the argument is complicated slightly by the appearance of the eigenvalue $\lambda$ in the boundary condition
\begin{equation} \label{SLBC2}
\xi\frac{{\rm d}\Phi_1}{{\rm d} \xi}+\scM_*k\Phi_1-\lambda(\scM_*-1)k^2\Phi_1=0
	\quad\text{at}\quad\xi=\xi_*	\,.
\end{equation}
However, if (\ref{SLBC2}) is replaced either by the boundary condition (i) $\xi\Phi_{1\xi}+\scM_*k\Phi_1=0$ or by (ii) $\Phi_1(\xi_*)=0$ then in each case we obtain a standard Sturm--Liouville problem with a discrete unbounded spectrum. It can be argued that the sequence of eigenvalues with condition (\ref{SLBC2}) is bracketed by the sequences with conditions (i) and (ii), and hence that the eigenvalues with (\ref{SLBC2}) also form a discrete unbounded spectrum; once again, for fixed $k$, $\sigma_n\to-1$ from above as $n\to\infty$.

\bibliographystyle{jfm}
\bibliography{references}

\end{document}